\documentclass[conference]{IEEEtran}

\usepackage{graphicx}

\begin{document}

\title{Method of Power Recycling in Co-Axial Mach Zender Interferometers for Low Noise Measurements}

\author{\authorblockN{Stephen Parker, Eugene Ivanov and Michael Tobar}
\authorblockA{School of Physics\\
University of Western Australia\\
Crawley, Western Australia 6009\\
Email: stephen.parker@physics.uwa.edu.au}}

\maketitle

\begin{abstract}
We present the first experimental study of a new type of power recycling microwave interferometer designed for low noise measurements. This system enhances sensitivity to phase fluctuations in a Device Under Test, independent of input power levels. The single sideband thermal white phase noise floor of the system has been lowered by 8 dB (reaching -185 dBc/Hz at 1 kHz offset frequency) at relatively low power levels (13 dBm).
\end{abstract}

\IEEEpeerreviewmaketitle

\section{Introduction}

Interferometers operating at microwave frequencies have been used for noise measurements for many years \cite{Whitwell}.  The quality of any measurement system is limited by its own noise properties. It is therefore beneficial to lower the noise floor of a measurement system as this will increase its quality. Several approaches have been taken in order to lower the noise floor of the interferometric measurement system. Cross correlation schemes allow significant improvements to noise properties but require lengthy data processing times \cite{Walls}.

In the 1980s Drever proposed a power recycling scheme to increase the sensitivity of an optical interferometer \cite{Drever}. The first recycling microwave interferometer was developed in 2002, which allowed real time measurements below the standard thermal noise limit \cite{Realtime}. The drawback of this system was that it could only be applied to bi-directional devices. In 2005 a true power recycling microwave interferometer scheme was proposed in an effort to increase the sensitivity of a Local Lorentz Invariance experiment \cite{LLI}. The work presented in this paper is the first study of this interferometric noise measurement system.

\section{Power Recycling}

\begin{figure} 
\centering
\includegraphics[width=0.5\textwidth]{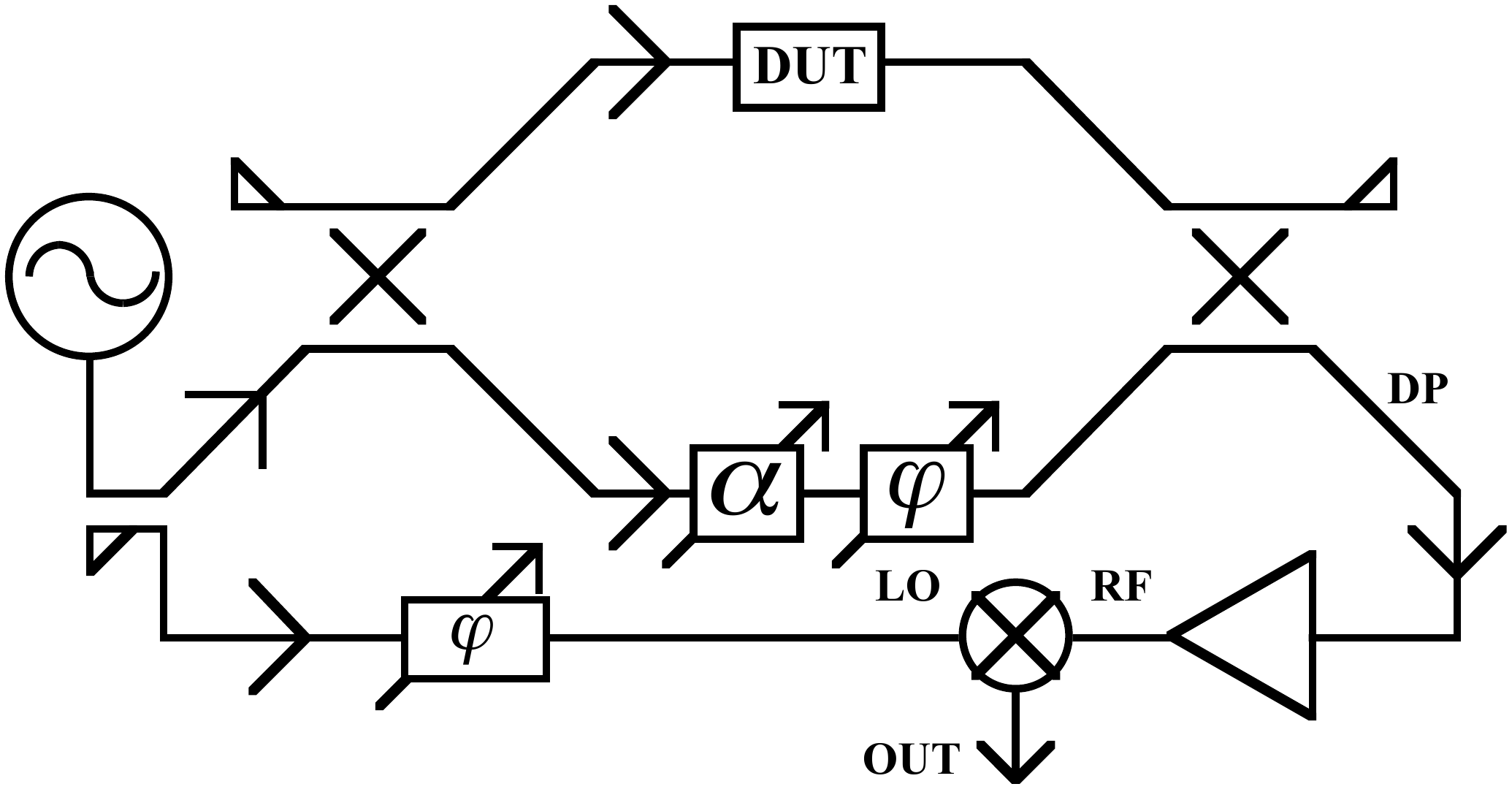}
\caption{\label{fig:circuitdiag1}A microwave interferometer with a phase / amplitude noise detection system.}
\end{figure}

The standard interferometric noise measurement system is shown in Fig. \ref{fig:circuitdiag1}. Noise fluctuations in the Device Under Test (DUT) are imprinted onto the carrier signal as sidebands offset from the carrier frequency. The compensating arm of the interferometer allows carrier suppression to be achieved, leaving just the noise sidebands, which are small in magnitude. The detection system uses a Low Noise Amplifier (LNA) to enhance these sidebands and overcome the technical noise of the non-linear mixer. The mixer outputs a voltage noise representing noise fluctuations in a DUT. Setting the phase of the signal at the LO port of the mixer to be in quadrature with the RF signal makes the system sensitive to phase fluctuations. The recorded voltage noise can then be converted into phase noise:
\begin{equation} \label{eq:volttophase}
S_{\varphi}(f) = \frac{S_{V}(f)}{S_{pd}^2} \frac{rad^{2}}{Hz}
\end{equation}
Where $S_{V}(f)$ is the spectral density of the voltage noise (given in $V^{2}$ / Hz) and $S_{pd}$ is the phase-to-voltage conversion efficiency of the interferometer. The contribution to the phase noise floor of this measurement system given by the detection electronics is known to be \cite{Expstudy}
\begin{equation} \label{eq:NFelec}
S_{\varphi}^{n/f}(f) = \frac{2 k_{B} T_{eff}}{P_{inc} \alpha_{DUT}}
\end{equation}
Where $k_{B}$ is the Boltzmann constant, $T_{eff}$ is the effective noise temperature of the detector system (comprised of the ambient temperature and the effective noise temperature of the LNA), $P_{inc}$ is the power incident on the DUT and $\alpha_{DUT}$ is the insertion loss (power ratio) of the DUT. 
\subsection{Limiting Noise}
We expect the detection electronics to be the primary limiting source of noise in our system \cite{Realtime}. In order to lower this limiting noise floor we can consider the following.

\subsubsection{Lowering the ambient temperature}
This is achievable through the use of cryogenics and would allow us to operate at 77 Kelvin (using liquid nitrogen) or 4 Kelvin (using liquid helium). Unfortunately the use of cryogenics complicates the situation, requiring components specifically designed to operate at cryogenic temperatures, bulky chambers to house the apparatus and ongoing running costs for the coolant used. In order to keep our system compact, simple and low cost we will proceed without the aid of cooling, knowing that the idea can be revisited.

\subsubsection{Using a better LNA}
We shall ensure that we are using the LNA with the best noise properties available to us.

\subsubsection{Decrease the insertion loss of the DUT}
Care must be taken to ensure that all cables and components are functioning properly, as these factors will contribute to higher attenuation in the system. The coaxial interfaces should be kept clean as metal dust particles exposed to high intensity electromagnetic fields can generate technical noise.  The insertion loss of the DUT can only be reduced so far.

\subsubsection{Increase the power incident on the DUT}
The power of the signal source can be increased within a reasonable level that will still allow carrier suppression to be maintable without the use of voltage controlled devices in a feedback loop. In addition to this, let us consider using the power available to us more efficiently - via \emph{power recycling}.

\subsection{Power Recycling}

\begin{figure} 
\centering
\includegraphics[width=0.5\textwidth]{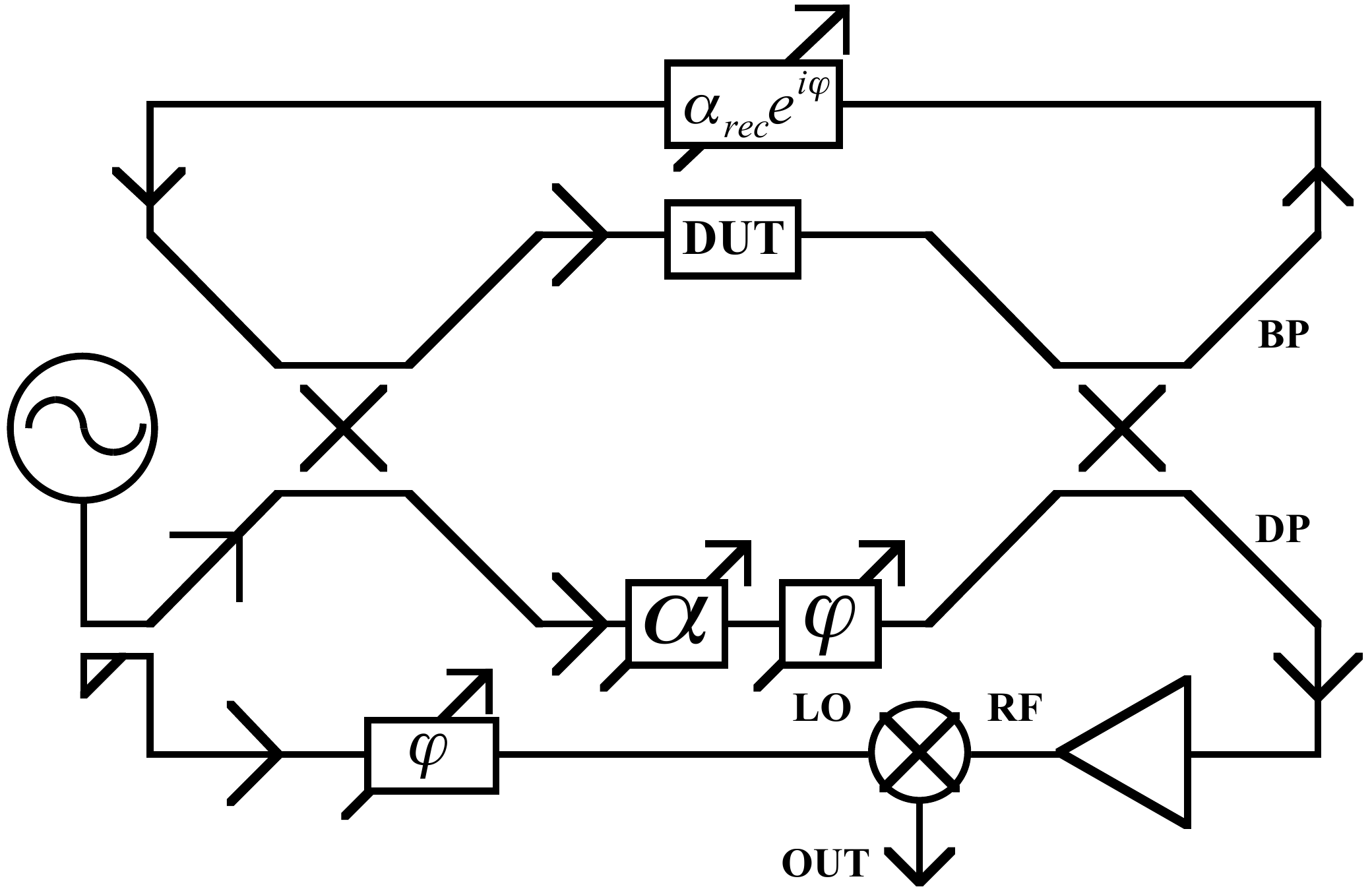}
\caption{\label{fig:circuitdiag2}A power recycling microwave interferometer with a phase / amplitude noise detection system.}
\end{figure}

Fig. \ref{fig:circuitdiag2} shows the simplified circuit diagram of a power recycling interferometric measurement system. The bright port of the interferometer (being the addition of the signal in the two arms) is taken and phase adjusted (so that the signal will constructively interfere with the signal at the input of the DUT) before being fed into the 3 dB hybrid used to initially split the signal.

\begin{figure} 
\centering
\includegraphics[width=0.48\textwidth]{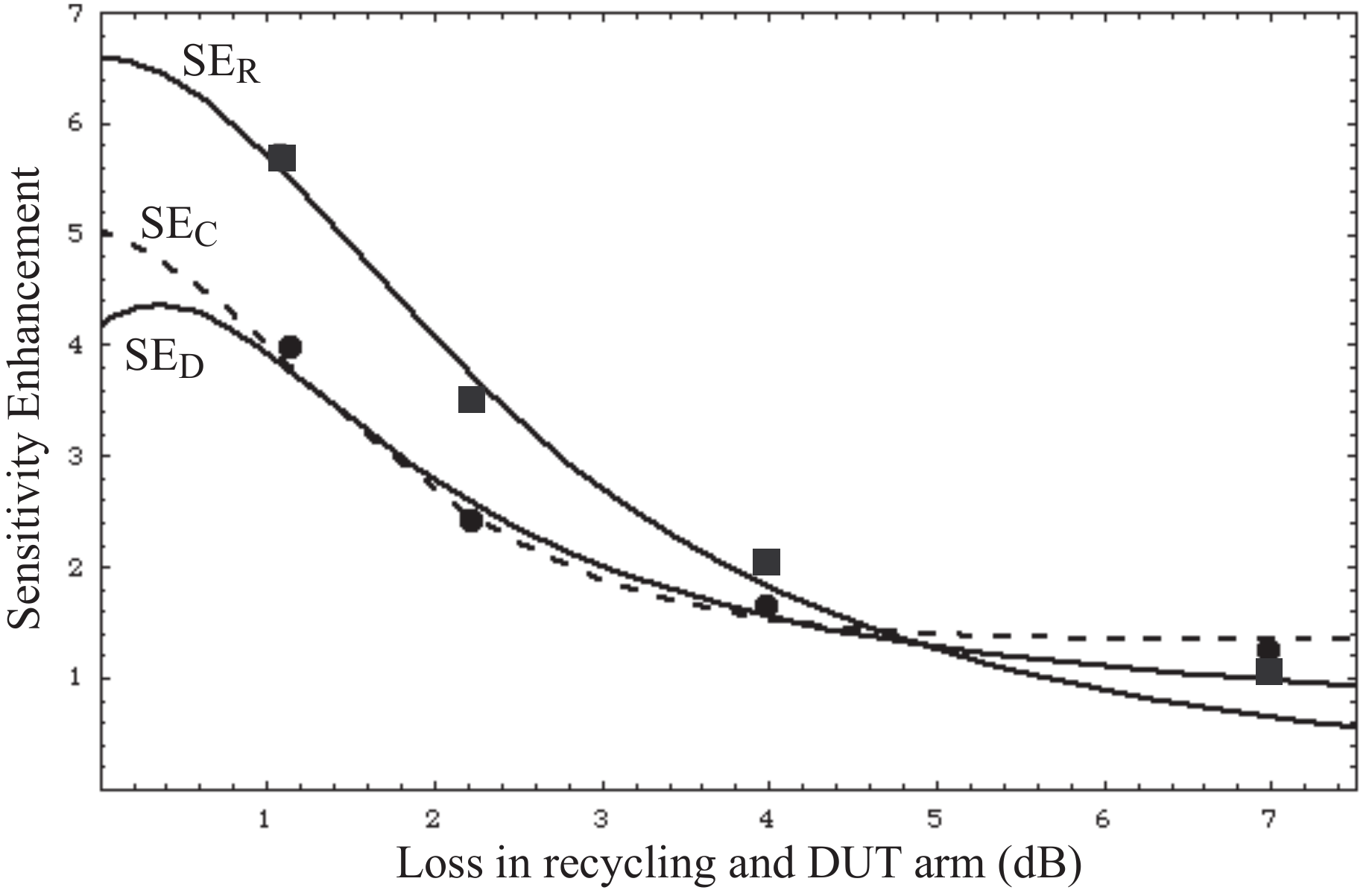}
\caption{\label{fig:SE}Sensitivity Enhancement as a function of loss in the DUT and recycling arms for the analytical functions ($SE_{D}$, $SE_{R}$ and $SE_{C}$ represent sensitivity to phase fluctuations in the DUT, recycling and compensation arms respectively) and numerical simulations (circular and square plot points for the DUT/compensation arms and recycling arm respectively).}
\end{figure}

\begin{figure}
\centering
\includegraphics[width=0.48\textwidth]{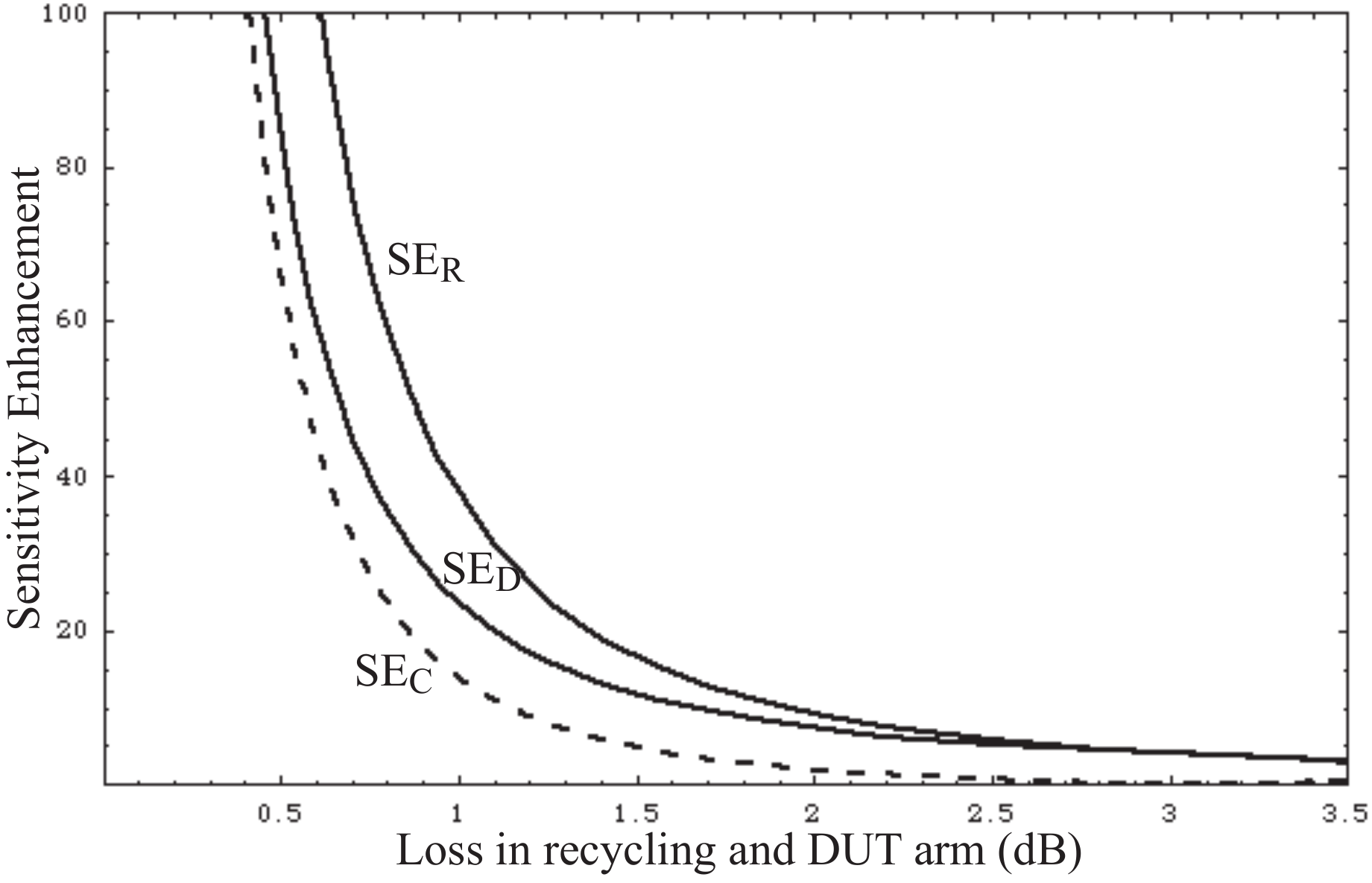}
\caption{\label{fig:SEideal}Ideal Sensitivity Enhancement as a function of loss in the DUT and recycling arms for the analytical functions ($SE_{D}$, $SE_{R}$ and $SE_{C}$ represent sensitivity to phase fluctuations in the DUT, recycling and compensation arms respectively).}
\end{figure}

To demonstrate the improvements that this system offers we shall consider the phase-to-voltage conversion efficiency ($S_{pd}$). Analysis of the ÒvoltageÓ in the circuit allows us to determine the $S_{pd}$ of this system:
\begin{equation} \label{eq:Spd}
S_{pd} = \frac{dV}{d\phi} \frac{Volts}{Radian}
\end{equation}
Sensitivity to phase fluctuations in the DUT, recycling and compensation arm will be considered. Comparing the $S_{pd}$ of the power recycling system to the $S_{pd}$ of the conventional system yields the Sensitivity Enhancement (SE) that the new scheme offers. For phase fluctuations in the DUT arm:
\begin{equation} \label{eq:SEdut}
SE_{D} = \frac{S^{PR}_{pd}}{S^{conv}_{pd}} = \frac{1}{(1 - \dot{r}\dot{T})^{2}}
\end{equation}
For phase fluctuations in the recycling arm:
\begin{equation} \label{eq:SErec}
SE_{R} = \frac{2\dot{r}\dot{T}}{(1 - \dot{r}\dot{T})^{2}}
\end{equation}
For phase fluctuations in the compensation arm:
\begin{equation} \label{eq:SEcomp}
SE_{C} = \frac{1 - 2\dot{r}\dot{T}}{(1 - \dot{r}\dot{T})^{2}}
\end{equation}
In all cases $\dot{r}$ and $\dot{T}$ represent complex transfer functions describing the attenuation and phase in the recycling arm and DUT respectively. The two attenuation factors together represent the total loss in the DUT and recycling arm (not including losses associated with the 3 dB hybrid directional couplers). In deriving equations \ref{eq:SEdut}, \ref{eq:SErec} and \ref{eq:SEcomp} it is assumed that the recycled signal has undergone an optimal phase shift to give the maximum possible constructive interference with the signal incident on the DUT and that the carrier is suppressed and the signal does not drift away from this point. Equations \ref{eq:SEdut}, \ref{eq:SErec} and \ref{eq:SEcomp} can be tested against numerical simulations by including an established model for the mixer \cite{Realtime}. Physical values are used to determine the parameters required for carrier suppression and then the mixer voltage output is plotted as a function of phase to find the $S_{pd}$. The analytical functions are scaled and phase adjusted correctly and plotted with the corresponding numerical values in Fig. \ref{fig:SE}. This figure indicates that these two methods are congruent and give confidence to the model. The system appears to be most sensitive to phase fluctuations in the recycling arm. Sensitivities to phase fluctuations in the DUT and compensation arms are identical for the numerical values. For the analytical functions they start to deviate for low and high losses. If the phase difference associated with the product $\dot{r}\dot{T}$ is set to 0 then the ideal case is obtained (shown in Fig. \ref{fig:SEideal}) giving the maximum Sensitivity Enhancement possible. It is apparent that the recycling loop loss parameter is very important and that it is beneficial to reduce this loss as much as possible. To keep the insertion loss of the recycling arm at a minimum, the phase shifter will be removed. In order to achieve carrier suppression the phase of the product $\dot{r}\dot{T}$ can be adjusted by changing the frequency of the pump oscillator. The frequency range needed for tuning will be on the order of hundreds of Megahertz and the properties (e.g. attenuation) of components being used in the circuit exhibit an insignificant frequency dependency over this range.

\section{Experimental Results and Discussion}

The circuit shown in figure \ref{fig:circuitdiag1} was constructed and tested before being adapted to create the power recycling circuit from figure \ref{fig:circuitdiag2}, which was then tested. In both cases the DUT was a short piece of cable which has minimal attenuation.  The pump oscillator produced a signal of frequency 10 GHz and power 13 dBm. 

\subsection{Conversion Efficiencies}

\begin{figure}
\centering
\includegraphics[width=0.48\textwidth]{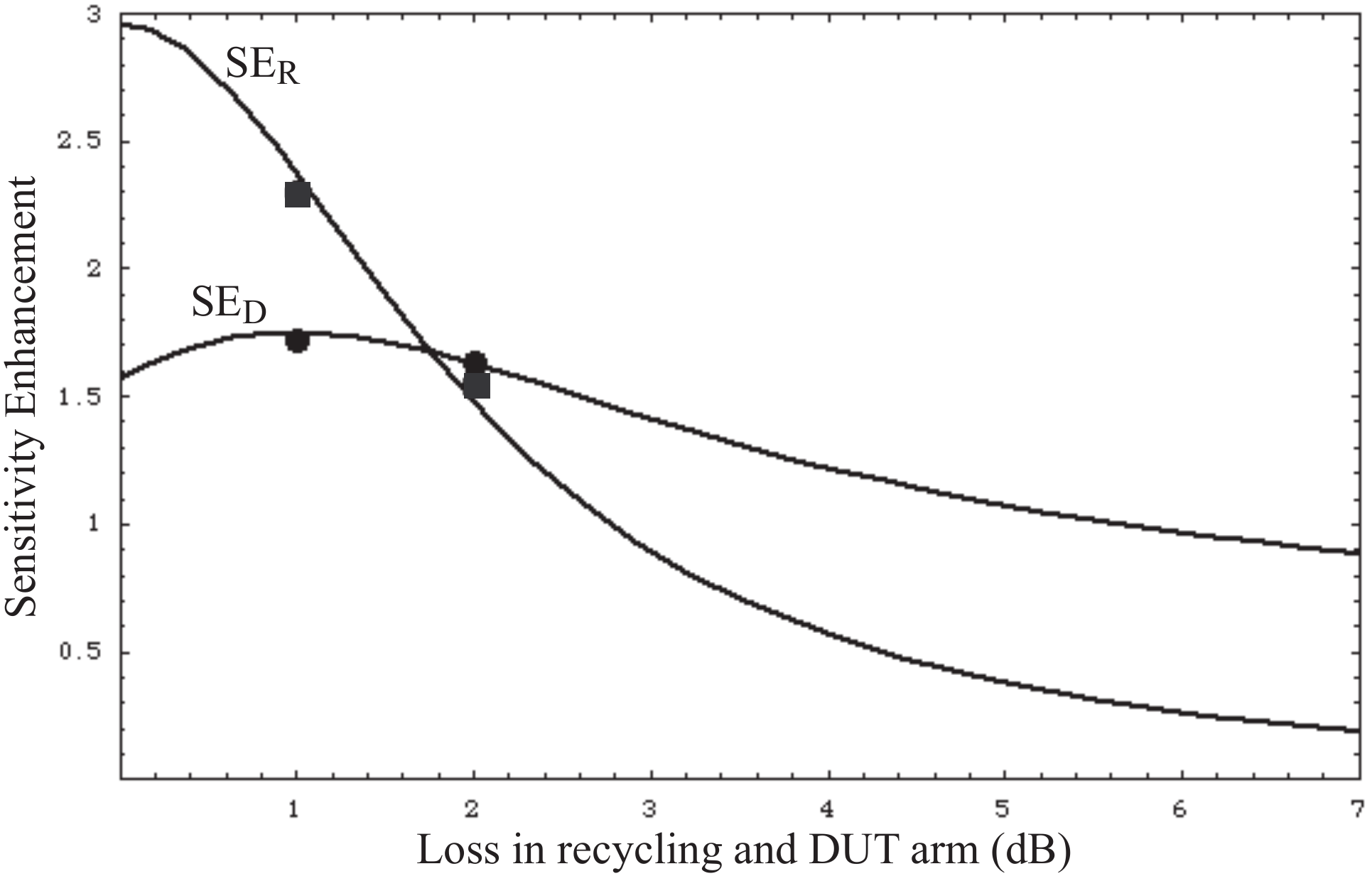}
\caption{\label{fig:SE_exp}Sensitivity Enhancement as a function of loss in the DUT and recycling arms for the analytical functions ($SE_{D}$ and $SE_{R}$ represent sensitivity to phase fluctuations in the DUT and recycling arms respectively). The circular and square plot points are experimental values for the DUT arm and recycling arm respectively.}
\end{figure}

The conversion efficiency of the system is measured by observing the change in mixer output voltage caused by the introduction of some phase mismatch in the interferometer. If the phases of the RF signal and the LO signal are in quadrature then the relationship between the mixer output voltage and the phase mismatch is linear around the point of carrier suppression. The gradient of this slope is the conversion efficiency of the system.  A voltage controlled phase shifter was introduced into the system to increase the accuracy and reproducibility of the the measurements. It was removed for the noise measurements. The $S_{pd}$ for the conventional system was measured to be 14 V/rad. The $S_{pd}$ of the power recycling system was measured to be 25 V/rad for the DUT arm and 34 V/rad for the recycling arm. This corresponds to a Sensitivity Enhancement of 1.8 and 2.4 respectively. To test the validity of our analytical model some extra losses were introduced into the system and these measurements repeated. Fig. \ref{fig:SE_exp} shows these results compared to the analytical plots (equations \ref{eq:SEdut} and \ref{eq:SErec}). Results for the compensation arm where the same as those for the DUT arm (as predicted by the analytical and numerical simulations). Both curves in figure \ref{fig:SE_exp} have been scaled and phase adjusted by different amounts in order to recreate the experimental values. The losses are due to devices such as phase shifters, cables and connectors. At very low losses we expect the sensitivity enhancement to be much higher than is currently predicted. It is assumed that in this regime the model is incomplete and needs to be further refined. As seen in figures \ref{fig:SE} and \ref{fig:SE_exp} it will be possible to obtain a lower noise floor by placing the DUT in the recycling arm, assuming that the loss in the DUT is small enough. We know that enhancing phase sensitivity by a factor of 2.4 will result in an 8 dB lowering of the phase noise floor of the system. If we were to measure the noise floor of a DUT with greater attenuation then the improvements would not be as significant.

\subsection{Noise Floors}

\begin{figure} 
\centering
\includegraphics[width=0.5\textwidth]{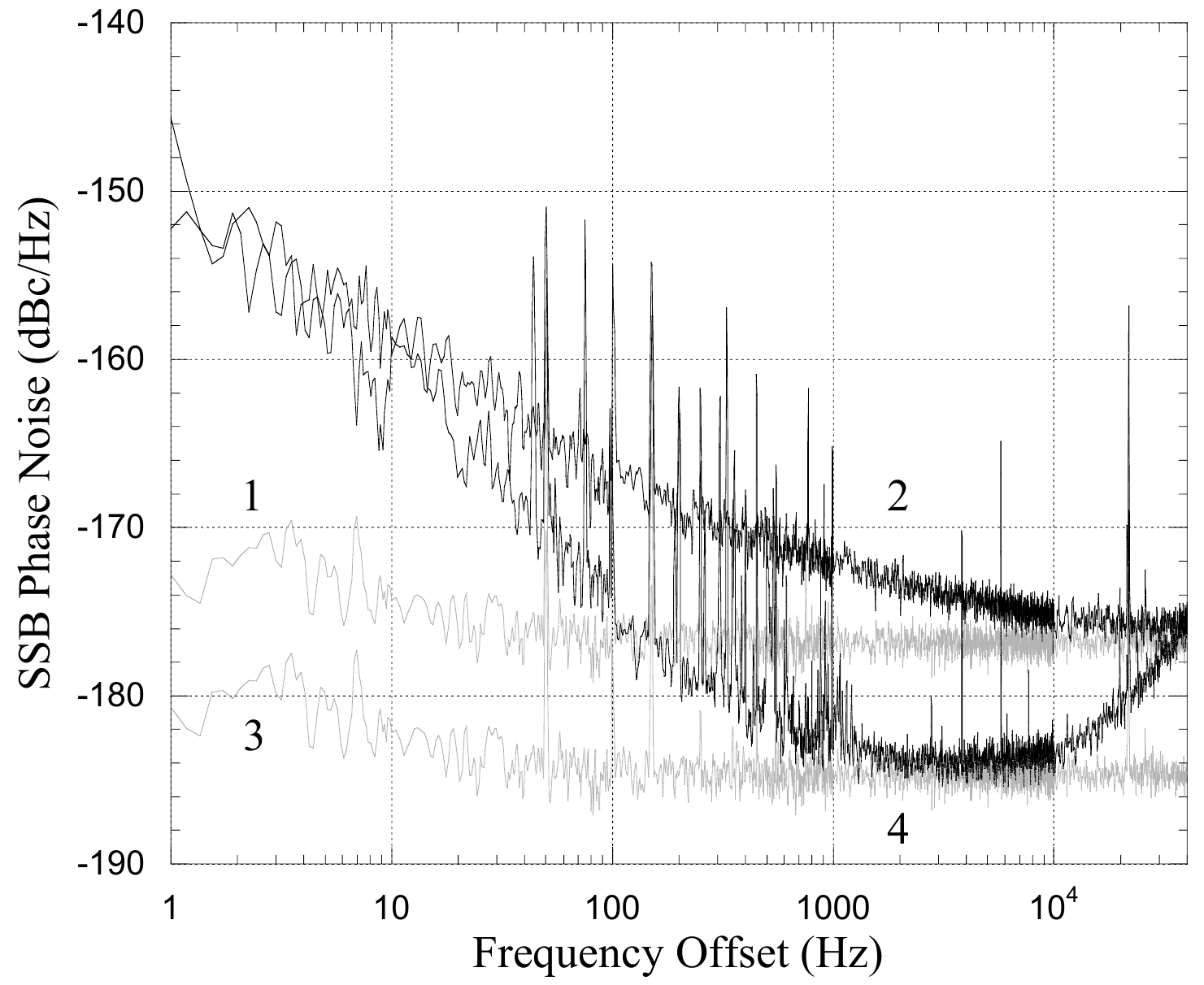}
\caption{\label{fig:NFs}Phase noise floor of the detection electronics for the conventional interferometer (curve 1), phase noise of the conventional interferometer (curve 2), phase noise floor of the detection electronics for the power recycling interferometer (curve 3) and phase noise of the power recycling interferometer (curve 4).}
\end{figure}

Fig. \ref{fig:NFs} shows the measured Singe Side Band (SSB) noise properties of the two interferometers compared to the noise limiting detection electronics. Being sensitive to phase changes in the recycling arm or DUT arm both result in the same noise properties for the power recycling interferometer, the only difference being a $S_{pd}$ dependant shifting of the curve up or down. The noise floor of the detection electronics was measured by terminating the input of the LNA and then recording the output of the mixer with a fast Fourier transform vector signal analyzer. At low offset frequencies both interferometers are dominated by 1/f flicker noise. This is a common observation but the exact source of this noise remains unknown, although it is theorized to arise from a number of factors including ambient temperature fluctuations.

At higher offset frequencies the noise plots of the two interferometers diverge. The peaks in the 100 Hz Ð 1 kHz window are attributed to vibrations and the formation of DC ground loops. From 10 kHz and above the conventional system becomes limited by the white noise floor of the detection electronics, reaching -177 dBc / Hz. This result is consistent with previous work \cite{Noisereduc}, where a white noise floor of -193 dBc / Hz was reached using an input power of approximately 20 dBm. From (\ref{eq:volttophase}), the 7 db difference in power levels corresponds to a 14 dB difference in white noise floors. The 2 dB discrepancy can be attributed to differences in losses and components used. The power recycling interferometer reaches the detection electronics noise floor (-185 dBc / Hz) at 1 kHz frequency offset. At an offset frequency of 10 kHz the noise of the power recycling interferometer begins to rise, following a ÒfÓ curve. This rise in noise is due to an increased sensitivity to oscillator phase noise.

\subsection{Oscillator Noise}

\begin{figure} 
\centering
\includegraphics[width=0.5\textwidth]{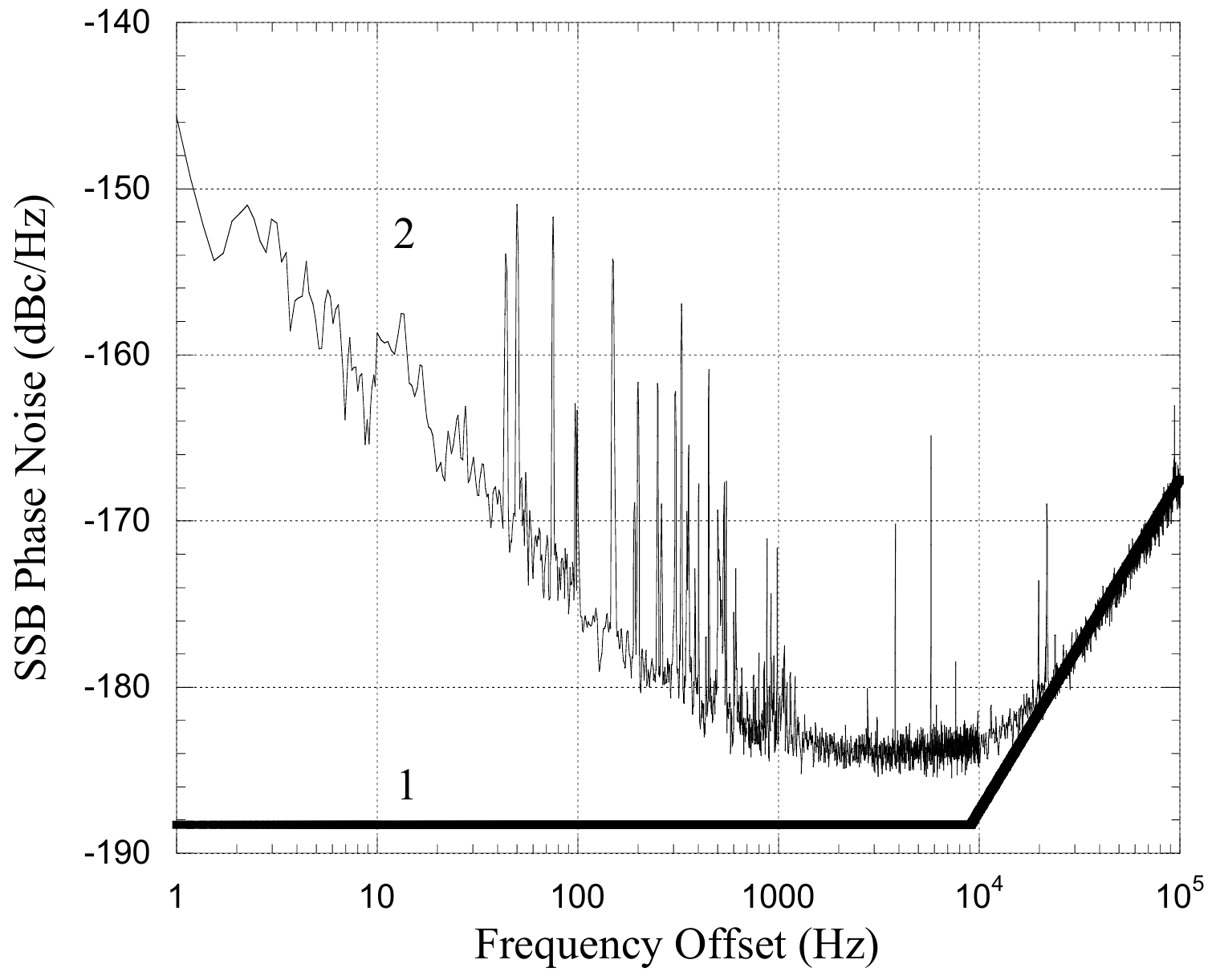}
\caption{\label{fig:oscil}Oscillator phase noise contribution calculated from (\ref{eq:NFoscill}) (curve 1) and the experimentally measured phase noise floor of the power recycling interferometer (curve 2).}
\end{figure}

The recycling arm of the interferometer will act as a dispersive element, due to its frequency dependance. The system will now have an increased sensitivity to frequency fluctuations of the pump oscillator. If we observe the power recycling circuit it is possible to determine the phase noise floor of the interferometer due to the oscillator phase fluctuations:
\begin{equation} \label{eq:NFoscill}
S^{n/f}_{\varphi}(f) \propto S^{osc}_{\varphi}(f) (f \tau_{rec})^{2} \alpha_{rec} \alpha_{DUT}
\end{equation}
Where $\tau_{rec}$ is the time delay in the recycling arm. Values for the phase noise of the pump oscillator (Agilent E8257C) are provided by the manufacturer \cite{Agilent}. Fig. \ref{fig:oscil} shows this calculated noise floor compared to the measured noise floor of the power recycling interferometer. It is clear that the high frequency noise is caused by the phase noise of the pump oscillator.

In order to eradicate this high frequency noise the phase noise of the oscillator needs to be lowered. The oscillator used for these experiments was not of high quality. An oscillator with less phase noise would not contribute to the noise floor of the interferometer.

\subsection{Improvements}

An improvement of 8 dB to the noise floor of the system has been demonstrated, however it should be possible to further improve upon this. Looking at figure 3 we note that our observed sensitivity enhancement of 2.4 is not reaching the full potential of the power recycling interferometer - reducing the losses in the recycling loop will yield greater enhancements. Consider the introduction of an amplifier and variable attenuator to the recycling arm. By effectively adjusting the gain of the LNA (through the use of the variable attenuator) it would be possible to compensate for the losses in the recycling arm and achieve a recycling loop loss of 0 dB. The overall gain in the recycling loop needs to be less than unity to avoid the development of oscillatory type instabilities. It is also possible that the phase noise of the amplifier would degrade the system noise floor enough so as to make the idea unfeasible. These effects have yet to be properly explored.

\section{Conclusion}
A power recycling microwave interferometer has been constructed. It is simple in design and implementation and increases the sensitivity to phase fluctuations compared to that of the standard interferometric noise measurement system at the same level of input power. An 8 dB lowering of the phase noise floor of a measurement system has been demonstrated, achieving -185 dBc/Hz at 1 kHz frequency offset for relatively low power levels (13 dBm). Further improvements appear to be achieveable.

\section*{Acknowledgment}

The authors would like to thank the ARC and PSI for support, and Jean-Michel LeFloch for the data acquisition software.

% http://www.ctan.org/tex-archive/macros/latex/contrib/supported/IEEEtran/bibtex
\bibliographystyle{IEEEtran.bst}
\bibliography{refs}

\end{document}